\documentclass[11pt]{article}
\usepackage{amsmath,amsfonts,latexsym,graphicx,amssymb}
\usepackage{amsfonts}
\usepackage{amssymb}
\usepackage{mathrsfs}
\usepackage{amsthm}
\usepackage{graphicx}
\usepackage{amsmath}
\date{}
\usepackage{amsfonts}
\usepackage{amssymb}
\usepackage{mathrsfs}
\usepackage{amsthm}
\usepackage{graphicx}
\usepackage{amsmath}

\newcommand{\<}{\langle}
\renewcommand{\>}{\rangle}

\newcommand{\ot}{{\,\otimes\,}}
\newcommand{{\Cd}}{{\mathbb{C}^d}}

\def\<{\langle}
\def\>{\rangle}

\newtheorem{theorem}{Theorem}

\begin{document}

\title{\bf {Optimal entanglement witnesses for two qutrits}} \author{Dariusz
Chru\'sci\'nski and Gniewomir Sarbicki \\
Institute of Physics, Nicolaus Copernicus University,\\
Grudzi\c{a}dzka 5/7, 87--100 Toru\'n, Poland}

\maketitle

\begin{abstract}
We provide a proof that  entanglement witnesses considered recently in \cite{Filip}
are optimal.
\end{abstract}

\maketitle


In a recent paper \cite{Filip} we analyzed z class of entanglement witnesses (EW) given by
\begin{equation}\label{}
  W[a,b,c]\, =\, \left( \begin{array}{ccc|ccc|ccc}
    a & \cdot & \cdot & \cdot & -1 & \cdot & \cdot & \cdot & -1 \\
    \cdot& b & \cdot & \cdot & \cdot& \cdot & \cdot & \cdot & \cdot  \\
    \cdot& \cdot & c & \cdot & \cdot & \cdot & \cdot & \cdot &\cdot   \\ \hline
    \cdot & \cdot & \cdot & c & \cdot & \cdot & \cdot & \cdot & \cdot \\
    -1 & \cdot & \cdot & \cdot & a & \cdot & \cdot & \cdot & -1  \\
    \cdot& \cdot & \cdot & \cdot & \cdot & b & \cdot & \cdot & \cdot  \\ \hline
    \cdot & \cdot & \cdot & \cdot& \cdot & \cdot & b & \cdot & \cdot \\
    \cdot& \cdot & \cdot & \cdot & \cdot& \cdot & \cdot & c & \cdot  \\
    -1 & \cdot& \cdot & \cdot & -1 & \cdot& \cdot & \cdot & a
     \end{array} \right)\ ,
\end{equation}
where to make the picture more transparent we replaced zeros by dots (for simplicity we skipped the normalization factor which is not essential). One proves the following result \cite{Cho-Kye}

\begin{theorem}  \label{TH-korea}
$W[a,b,c]$ defines an entanglement witness if and only if
\begin{enumerate}
\item $0 \leq a < 2\ $,
\item $ a+b+c \geq 2\ $,
\item if $a \leq 1\ $, then $ \ bc \geq (1-a)^2$.
\end{enumerate}
Moreover, being EW it is indecomposable if and only if $bc < (2-a)^2/4$.
\end{theorem}
In particular we analyzed \cite{Filip} a subclass of EWs defined by
\begin{equation}\label{abc}
0 \leq a \leq 1\ , \ \ \ a+b+c=2\ , \ \ \ bc = (1-a)^2\ .
\end{equation}
The corresponding EWs $W[b,c]:=W[2-b-c,b,c]$ belong to the ellipse on $bc$-plane -- see Fig.~1.
\begin{figure}[htp] \label{fig}
 \centering
\includegraphics[scale=0.65]{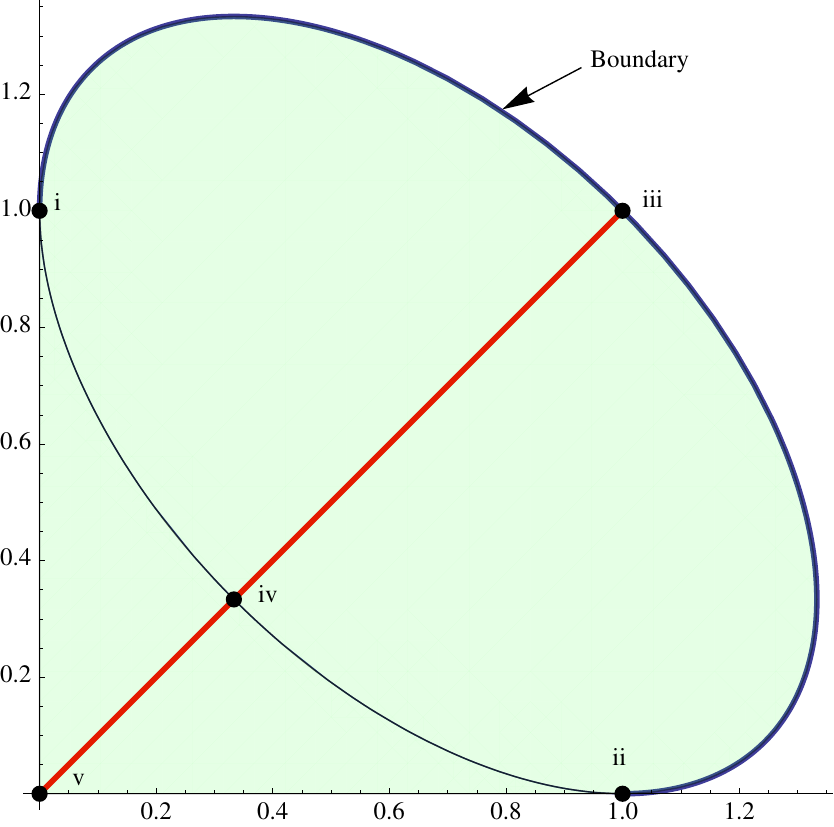}
\caption{A convex set of EWs $W[b,c]$. A line $b=c$ corresponds to decomposable EW. Special points: (i) and (ii) Choi EWs, (iii) EW corresponding to reduction map, (v) positive operator with $b=c=0$, (iv) decomposable EW with $b=c=1/3$.}
\end{figure}
 It was conjectured \cite{Filip} that $W[b,c]$ are optimal. In the present paper we show that this conjecture is true.
\begin{theorem}
EWs $W[b,c]$ defined by (\ref{abc}) are optimal.
\end{theorem}

\noindent Proof: let us define
\begin{equation}\label{}
    \mathcal{P}_{bc} = \{ x \ot y \in \mathbb{C}^3 \ot \mathbb{C}^3\ | \ \< x\ot y|W[b,c]|x \ot y\> =0  \} \ .
\end{equation}
It is well known \cite{Lew} that if the set $\mathcal{P}_{bc}$ spans the entire Hilbert
space $\mathbb{C}^3 \ot \mathbb{C}^3$, then $W[b,c]$ is an optimal EW. If we find a set of vectors  $y \in \mathbb{C}^3$  such that the $3 \times 3$ matrix
\begin{equation}\label{}
    W_y[b,c] := {\rm Tr}_2( W[b,c] \cdot \mathbb{I}_3 \ot |y\>\<y|)\ ,
\end{equation}
is singular, then for each vector $x_y$ belonging to the kernel of $W_y[b,c]$ the product vector $x_y \ot y$ belongs to $\mathcal{P}_{bc}$ (${\rm Tr}_2$ denotes a partial trace over the second factor in $\mathbb{C}^3 \ot \mathbb{C}^3$).
 The matrix $W_y[b,c]$ is given by the formula
 \begin{align}
  & W_y[b,c]=\left[ \begin{array}{ccc} a|y_1|^2 + b|y_2|^2 + c|y_3|^2 & y_1^* y_2 & y_1^* y_3 \\ y_2^* y_1 & c|y_1|^2 + a|y_2|^2 + b|y_3|^2 & y_2^* y_3 \\ y_3^* y_1 & y_3^* y_2 & b|y_1|^2 + c|y_2|^2 + a|y_3|^2 \end{array} \right] \label{W_bc}\nonumber  \\
  & = \left[ \begin{array}{ccc} (a+1)|y_1|^2 + b|y_2|^2 + c|y_3|^2 & 0 & 0 \\ 0 & c|y_1|^2 + (a+1)|y_2|^2 + b|y_3|^2 & 0 \\ 0 & 0 & b|y_1|^2 + c|y_2|^2 + (a+1)|y_3|^2 \end{array} \right] - |y^* \rangle \langle y^* | \nonumber
 \end{align}
Let us observe, that for any $a,b,c$ satisfying Theorem \ref{TH-korea} and $y=[e^{i \alpha},e^{i \beta},e^{i \gamma}]$ one finds
 \begin{displaymath}
W_y[b,c]=  \mathrm{diag}[e^{-i\alpha},e^{-i\beta},e^{-i\gamma}]
  \left[ \begin{array}{ccc} 2 & -1 & -1 \\ -1 & 2 & -1 \\ -1 & -1 & 2 \end{array} \right]
  \mathrm{diag}[e^{i\alpha},e^{i\beta},e^{i\gamma}]  \ .
 \end{displaymath}
This matrix has rank $2$ and its 1-dim. kernel is spanned by the vector $x_y=[e^{-i \alpha},e^{-i \beta},e^{-i \gamma}]$. Hence we have the following continuous family of product vectors
 \begin{equation}
  x_y \otimes y = [1, e^{i(\beta-\alpha)}, e^{i(\gamma-\alpha)}, e^{i(\alpha-\beta)}, 1, e^{i(\gamma-\beta)}, e^{i(\alpha-\gamma)}, e^{i(\beta-\gamma)},1]
  \label{7-family}
 \end{equation}
Note that this family  spans at most 7-dimensional subspace of $\mathbb{C}^3 \otimes \mathbb{C}^3$. To show, that this subspace is exactly 7-dimensional, it suffices to consider the following set of $(\alpha,\beta,\gamma)$
 \begin{equation}  \label{SET}
  (0,0,0), \quad (0,0,\pi), \quad (0,\pi,0), \quad (0,\pi,\pi), \quad (0,0,\pi/2), \quad (0,\pi/2,0), \quad (0,\pi/2,-\pi/2)\ .
 \end{equation}
Consider now $y =(0,y_2,y_3)$. One has
 \begin{displaymath}
W_y[b,c] =  \left[ \begin{array}{ccc}
  b|y_2|^2 + c|y_3|^2 & 0 & 0 \\ 0 & a|y_2|^2 + b|y_3|^2 & - y_2^* y_3 \\ 0 & - y_3^* y_2 & c|y_2|^2 + a|y_3|^2\
 \end{array} \right]\ .
 \end{displaymath}
Its determinant is given by the formula:
 \begin{displaymath}
  \det W_y[b,c] = (b|y_2|^2 + c|y_3|^2)(ab|y_2|^4+(a^2+ac-1)|y_2|^2 |y_3^2|+bc|y_3|^4)\ .
 \end{displaymath}
We are looking for $y\in \mathbb{C}^3$, that the determinant vanishes.

\begin{center} Case 1: $b,c\ne 0$. \end{center}

\noindent Now, the first term is always positive and so the second term has to vanish. Taking $||y||=1$, one can replace $|y_3|^2$ by $1-|y_2|^2$. The second term reads as follows
 \begin{equation}
  a\,(4-3a)|y_2|^4+2\, a\,(a-b-1)|y_2|^2+ab = 0\ .
  \label{second_term}
 \end{equation}
 We use here relations $bc=(a-1)^2$ and $a=2-b+c$. One  also assume that $b<c$ (the case $c<b$ may be treated in the same way using a symmetry $b \longleftrightarrow c$ \cite{Filip}). One obtains the following formulae for $b$ and $c$
 \begin{align*}
  b=\frac 12 (2-a-\sqrt{4a-3a^2}\,)\ , \ \ \
  c=\frac 12 (2-a+\sqrt{4a-3a^2}\,)
 \end{align*}
 The discriminant of the quadratic equation (for $|y_2|^2$) vanishes (it can not be positive due to the fact that $W[b,c]$ is an EW) and one easily solves (\ref{second_term}) to get
 \begin{displaymath}
  |y_2|^2 =\frac{1+b-a}{4-3a}\ .
 \end{displaymath}
 The vector $y$ is then equal (after calculating $|y_3|^2$, we drop the normalization):
 \begin{equation}
  y=[0,\sqrt{1+b-a},\sqrt{3-b-2a}e^{i \phi}] =: [0,p,q e^{i\phi_1}]\ .
  \label{y}
 \end{equation}
 For such  $y$, the kernel of $W_y[b,c]$ is spanned by the vector
 \begin{equation}
  x_y=[0,y_2^* \cdot y_3, a|y_2|^2+b|y_3|^2] =: [0,r e^{i\phi_1},s]
  \label{x}
 \end{equation}
 The numbers $p,q,r,s$ are nonzero and depend only on parameters $a,b,c$. Let
 \begin{align*}
  & \Psi^{(1)} := x_y \ot y = [0,0,0,0,pr e^{i\phi_1}, ps, 0, qr e^{2i\phi_1}, qs e^{i\phi_1}] \ .
 \end{align*}
 Because of the cyclic symmetry of the problem, one can find the similar product vectors for $y_2=0$ and $y_3=0$:
 \begin{align*}
  & \Psi^{(2)}=[qs e^{i\phi_2},0,qr e^{2i\phi_2},0,0,0,ps, 0, pr e^{i\phi_2}] \ , \\
  & \Psi^{(3)}=[pr e^{i\phi_3}, ps, 0, qr e^{2i\phi_3},qs e^{i\phi_3},0,0,0,0]\ .
 \end{align*}
  Now, it turns out that 7 vectors from the family (\ref{7-family}) generated by a set (\ref{SET}) plus two arbitrary vectors from the family $(\Psi^{(1)},\Psi^{(2)},\Psi^{(3)})$ defines a basis in $\mathbb{C}^3 \ot \mathbb{C}^3$. Indeed, taking 7 vectors from (\ref{7-family}) and $\Psi^{(1)}$, $\Psi^{(2)}$ one obtains the following $9 \times 9$ matrix:
\begin{equation}\label{BIG}
    \left[ \begin{array}{ccccccccc}
  1& 1& 1& 1& 1& 1& 1& 1& 1 \\ 1& 1& -1& 1& 1& -1& -1& -1& 1 \\ 1& -1&
   1& -1& 1& -1& 1& -1& 1 \\ 1& -1& -1& -1& 1& -1& -1& 1& 1 \\ 1& 1&
  i & 1& 1&  i & -i & -i & 1 \\ 1& i & 1& -i & 1& -i & 1&  i& 1 \\1&
  i & -i & -i & 1& -1& i & i & 1 \\ 0& 0& 0& 0& pr e^{i\phi_1} & ps & 0&
  qr e^{2i\phi_1}& qs e^{i \phi_1} \\ qs e^{i \phi_2} & 0 & qr e^{2i\phi_2} & 0& 0& 0& ps& 0&
  pr e^{i\phi_2} \end{array} \right]\ .
\end{equation}
Its determinant reads
$$ (-32 + 160 i) e^{i(\phi_1+\phi_2)} [\, (qs)^2 + (pr)^2 - qspr\, ]  \ , $$
and is different from zero except $qs = pr=0$. Note, however, that for $b,c \neq 0$  one has $qs,pr \neq 0$.

\begin{center} Case 2: $b=0, c=1$. \end{center}

\noindent Now, the determinant reads
 \begin{displaymath}
 \det W_y[b,c] =  |y_1|^2 |y_2|^4 + |y_2|^2 |y_3|^4 + |y_3|^2 |y_1|^4 - 3 |y_1|^2 |y_2|^2 |y_3|^2\ .
 \end{displaymath}
 If one of coordinates, say $y_1$ is zero, then the determinant is equal $|y_2|^2 |y_3|^4$ and vanishes only if $y_2$ or $y_3$ vanishes, so the only vectors $y$ with at least one zero coordinate for which $W_y[b,c]$ vanishes are
 \begin{equation}\label{Phi}
    \Phi^{(1)} :=  [1,0,0]\otimes[0,0,1]\ , \ \  \Phi^{(2)} := [0,1,0]\otimes[1,0,0]\ , \ \   \Phi^{(3)} :=[0,0,1]\otimes[0,1,0]\ .
 \end{equation}
 Now we will look for the remaining  vectors and we  assume that all coordinates are non-zero. Dividing the determinant by $|y_1|^2 |y_2|^2 |y_3|^2$ and gets the following equation
 \begin{displaymath}
  \frac{|y_2|}{|y_3|}+\frac{|y_3|}{|y_1|}+\frac{|y_1|}{|y_2|} - 3 = 0\ .
 \end{displaymath}
Its LHS is nonnegative and vanishes only for $|y_1|=|y_2|=|y_3|$, and  hence
$$ y=[e^{i\alpha},e^{i\beta},e^{i\gamma}]\ ,\ \ \  x_y=[e^{-i\alpha},e^{-i\beta},e^{-i\gamma}]\ , $$
and one gets again the 7-dimensional family of vectors (\ref{7-family}). However, vectors $\Phi^{(k)}$ are not linearly independent from (\ref{7-family}). Therefore, $\mathcal{P}_{01}$ spans only 7-dim. subspace of $\mathbb{C}^3 \ot \mathbb{C}^3$.

Actually, one obtains $\Phi^{(k)}$ from $\Psi^{(k)}$ in the limit $b \to 0$. Let us recall that the determinant of (\ref{BIG}) vanishes only when $qs = pr=0$. Now, $p=s=0$ when $b=0$ and $c=1$, whereas $q=r=0$ when $b=1$ and $c=0$. Hence, apart from two witnesses corresponding to Choi maps $W[1,0]$ and $W[0,1]$, the remaining EWs have spanning property, i.e. $\mathcal{P}_{bc}$ spans $\mathbb{C}^3 \ot \mathbb{C}^3$,   and hence they are optimal. \hfill $\Box$

\vspace{.5cm}

As this paper was completed we were informed by professors Kil-Chan Ha and Seung-Hyeok Kye that they provided an independent proof of optimality \cite{Kye-nasza}. Moreover, they proved  \cite{Kye} that all witnesses $W[b,c]$ are exposed (and hence extremal) except  $W[1,1]$, $W[1,0]$ and $W[0,1]$.

\end{document}